\documentclass[twocolumn,showpacs,preprintnumbers,prl,fleqn,superscriptaddress]{revtex4-2}
\usepackage[english]{babel}
\usepackage[utf8]{inputenc}
\usepackage{SIunits}
\usepackage{graphicx}
\usepackage{dcolumn}
\usepackage{amsmath}
\usepackage{natbib}
\usepackage{bm}
\usepackage{textcomp}
\usepackage{color}
\usepackage{float}
\usepackage{booktabs}
\usepackage{multirow}
\usepackage{tabularx}
\usepackage{adjustbox}
\usepackage{hyperref}
\usepackage{tikz}
\usepackage{makecell}

\begin{document}

\title{Magneto-optical sensing of the pressure driven magnetic ground states in bulk CrSBr}

\author{Amit Pawbake}
\email{amit.pawbake@lncmi.cnrs.fr}\affiliation{LNCMI, UPR 3228, CNRS, EMFL, Universit\'e Grenoble Alpes, 38000 Grenoble, France}
\author{Thomas Pelini}
\affiliation{LNCMI, UPR 3228, CNRS, EMFL, Universit\'e Grenoble Alpes, 38000 Grenoble, France}
\author{Ivan Mohelsky}
\affiliation{LNCMI, UPR 3228, CNRS, EMFL, Universit\'e Grenoble Alpes, 38000 Grenoble, France}
\author{Dipankar Jana}
\affiliation{LNCMI, UPR 3228, CNRS, EMFL, Universit\'e Grenoble Alpes, 38000 Grenoble, France}
\author{Ivan Breslavetz}
\affiliation{LNCMI, UPR 3228, CNRS, EMFL, Universit\'e Grenoble Alpes, 38000 Grenoble, France}
\author{Chang-Woo Cho}
\affiliation{LNCMI, UPR 3228, CNRS, EMFL, Universit\'e Grenoble Alpes, 38000 Grenoble, France}
\author{Milan Orlita}
\affiliation{LNCMI, UPR 3228, CNRS, EMFL, Universit\'e Grenoble Alpes, 38000 Grenoble, France}
\author{Marek Potemski}
\affiliation{LNCMI, UPR 3228, CNRS, EMFL, Universit\'e Grenoble Alpes, 38000 Grenoble, France}
\affiliation{CENTERA Labs, Institute of High Pressure Physics, PAS, 01 - 142 Warsaw, Poland}
\author{Marie-Aude Measson}
\affiliation{Institut Neel, Universit\'e Grenoble Alpes, 38000 Grenoble, France}
\author{Nathan Wilson}
\affiliation{Walter Schottky Institut, Physics Department and MCQST, Technische Universitat
Munchen, 85748 Garching, Germany}
\author{Kseniia Mosina}
\affiliation{Department of Inorganic Chemistry, University of Chemistry and Technology Prague, Technicka 5, 166 28 Prague 6, Czech Republic}
\author{Aljoscha Soll}
\affiliation{Department of Inorganic Chemistry, University of Chemistry and Technology Prague, Technicka 5, 166 28 Prague 6, Czech Republic}
\author{Zdenek Sofer}
\affiliation{Department of Inorganic Chemistry, University of Chemistry and Technology Prague, Technicka 5, 166 28 Prague 6, Czech Republic}
\author{Benjamin A. Piot}
\affiliation{LNCMI, UPR 3228, CNRS, EMFL, Universit\'e Grenoble Alpes, 38000 Grenoble, France}
\author{M. E. Zhitomirsky}
\affiliation{Universit\'e Grenoble Alpes, CEA, Grenoble INP, IRIG, Pheliqs, 38000 Grenoble, France}
\affiliation{Institut Laue-Langevin,  F-38042 Grenoble Cedex 9, France}
\author{Clement Faugeras}
\email{clement.faugeras@lncmi.cnrs.fr} \affiliation{LNCMI, UPR 3228, CNRS, EMFL, Universit\'e Grenoble Alpes, 38000 Grenoble, France}

\date{\today }

\begin{abstract}
 Competition between exchange interactions and magnetocrystalline anisotropy may bring new magnetic states that are of great current interest. An applied hydrostatic pressure can further be used to tune their balance. In this work we investigate the magnetization process of a biaxial antiferromagnet in an external magnetic field applied along the easy axis. We find that the single metamagnetic transition of the Ising type observed in this material under ambient pressure transforms under hydrostatic pressure into two transitions, a first-order spin flop transition followed by a second order transition towards a polarized ferromagnetic state near saturation. This reversible tuning into a new magnetic phase is obtained in layered bulk CrSBr at low temperature by varying the interlayer distance using high hydrostatic pressure, which efficiently acts on the interlayer magnetic exchange, and is probed by magneto-optical spectroscopy.
\end{abstract}

\maketitle

The generation of new magnetic phases by an external manipulation of the magnetic state is a requirement for the implementation of new spintronic and of magnontronic functionalities. Successful manipulations of magnetic states have been achieved by using spin-polarized electrical currents~\cite{Katine2000}, electrostatic gating~\cite{Weisheit2007}, or ultra fast laser excitations~\cite{Beaurepaire1996,Nemec2018}. An alternative approach relies on the direct modification of the magnetic interactions by changing the lattice parameters and/or the super-exchange paths by applying strain~\cite{Cenker2022} or hydrostatic pressure~\cite{Li2019,Song2019}. Such profound modifications affect both direct exchange magnetic interaction, and spin-orbit interaction related effects such as magnetocrystalline anisotropy, which determines the preferential directions of spins in a magnetically ordered solid and the spin-waves energy.

Magnetic van der Waals (vdW) materials are layered crystals with magnetic properties and that can be thinned down to the monolayer limit. They exhibit a broad variety of ordered magnetic structures~\cite{Jiang2021,Wang2022}. Recently, vdW magnets have attracted a lot of interest and stimulated many experimental and theoretical studies aiming at understanding their magnetic properties and at using them within functional vdW heterostructures~\cite{Zhong2017,Ciorciaro2020}. The observation of long range magnetic order in the 2D limit~\cite{Huang2017,Gong2017} that is precluded by the Mermin-Wagner theorem~\cite{Mermin1966}, is understood as a consequence of substantial magnetocrystalline anisotropy and of the associated energy gap in the spin-wave dispersion~\cite{Jenkins2022}. vdW materials offer new exciting possibilities to engineer magnetic properties of solid state systems as they are characterized by a very strong structural anisotropy: individual layers are bound together by weak vdW interactions. As a result, the interlayer spacing, the vdW gap, can be tuned efficiently through the application of hydrostatic pressure~\cite{Yankowitz2019,Li2019,Song2019,Wei2022}. Such changes strongly affect the short range interlayer interactions~\cite{Xia2020,Pawbake2022}, and the competition between magnetic exchange interactions and magnetic anisotropies. New magnetic phases can be engineered in a magnetic system by changing the competition between anisotropies and exchange interactions.

\begin{figure}
\includegraphics[width=1\linewidth,angle=0,clip]{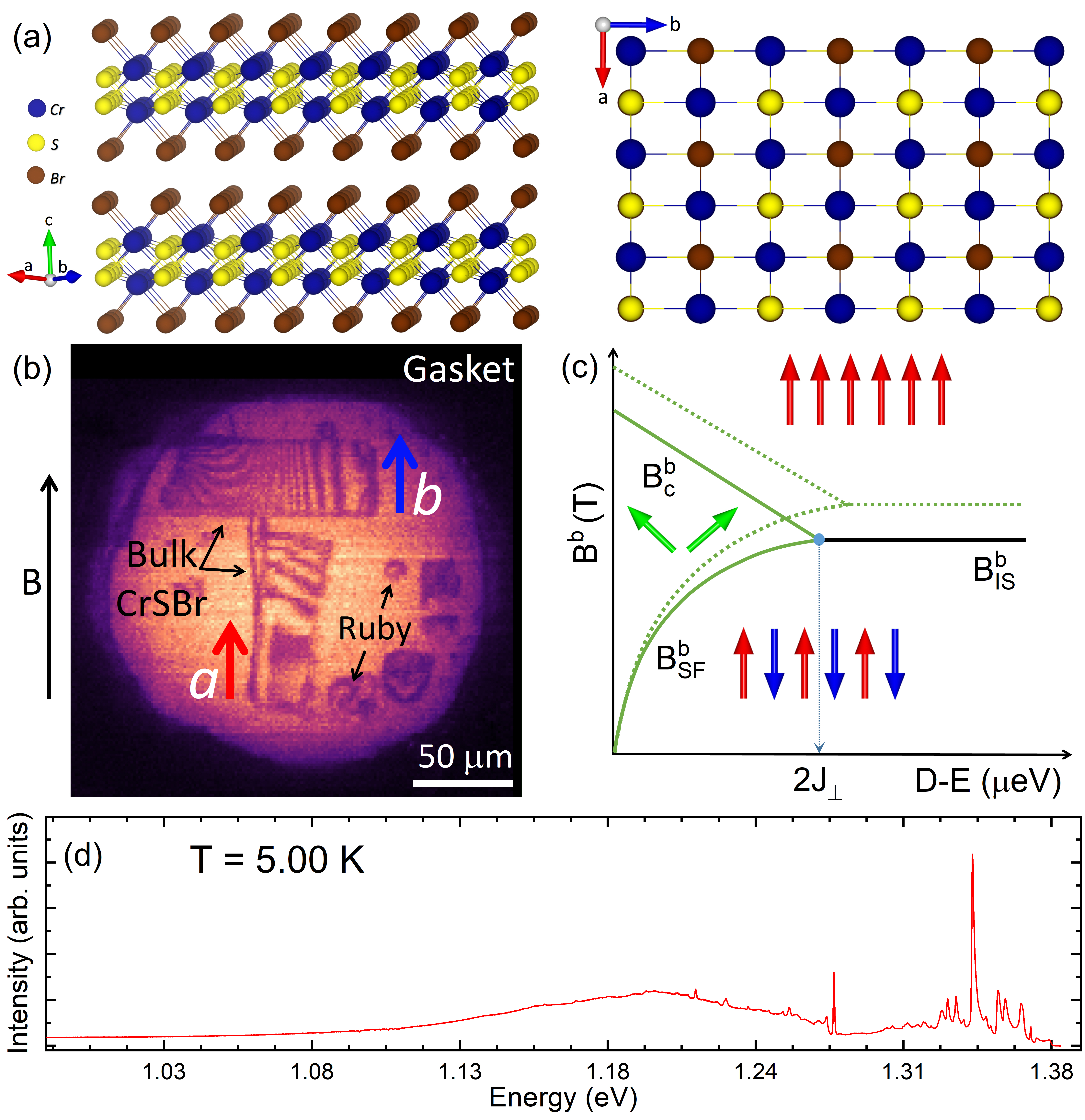}
\caption{a) Crystallographic structure of bulk CrSBr. b) Optical image of the pressure chamber of the diamond anvil cell showing two pieces of CrSBr with different orientations and ruby balls. c) (Solid lines) Schematics of the magnetic phase diagram of easy-plane AFM for a fixed value of $J_{\perp}$ showing the antiferromagnetic, the ferromagnetic and the spin-flop magnetic phases. The spin-flop phase appears when the anisotropy difference (D-E) becomes smaller than the interlayer exchange $J_{\perp}$. We indicate the critical values of magnetic field for spin-flip $B_c^b$, for the abrupt Ising transition $B^b_{IS}$ and for the spin-flop $B^b_{SF}$. The dashed lines represent a similar phase diagram with a higher value of $J_{\perp}$. d) Typical photoluminescence spectrum of bulk CrSBr at $T=5$~K showing the excitonic lines around $1.35$~eV and a broad emission at lower energies.
\label{Fig1}}
\end{figure}

In this Letter, we apply hydrostatic pressure to bulk CrSBr to tune the interlayer magnetic exchange interaction and the magnetocrystalline anisotropies. We sense the induced changes by magneto-photoluminescence measurements with an external magnetic field applied along one of the three crystallographic axis. The interlayer exchange parameter J$_{\perp}$ responsible for the antiferromagnetic (AFM) ordering is small in bulk CrSBr because of its lamellar structure. When decreasing the vdW gap, we observe an increase of J$_{\perp}$ by $1700~\%$ at $P>6$~GPa. This large increase with respect to magnetocrystalline anisotropies, stabilizes CrSBr into a particular magnetic ground state, characterized by an intermediate spin-flop phase appearing between the AFM ground state at $B=0$ and the ferromagnetic (FM) ground state when an external magnetic field, exceeding the saturation field, is applied (see Fig.~\ref{Fig1}b). This new magnetic ground state is observed because of the possibility to tune the competition between the interlayer exchange interaction and the magnetocrystaline anisotropies without changing the crystallographic stacking. Increasing further hydrostatic pressure, the in-plane magnetocrystalline anisotropy vanishes, transforming CrSBr into an easy-plane AFM.

CrSBr is an air-insensitive layered direct band gap semiconductor with E$_g=1.5$~eV and hosting tightly bound excitons that give rise to photoluminescence signals close to $1.35$~eV~\cite{Telford2020} at low temperature. It crystalizes in the P\textit{mmn} space group in an orthorhombic structure in which monolayers are bound together by vdW interactions, see Fig.~\ref{Fig1}a. From the point of view of magnetism, CrSBr belongs to the class of easy-axis layered AFM systems, with an in-plane anisotropy. Below the Curie temperature of $T_C=160$~K, magnetic moments in the layers orient ferromagnetically along the easy axis~\cite{Liu2022}, the crystallographic $\mathbf{b}$ axis. Below the N\'{e}el temperature of T$_N=132$~K, an AFM interaction couples each layer. Below $T=40$~K, various experimental signatures of a change in the magnetic ground state have been reported~\cite{Telford2020,Telford2021,LopezPaz2022,Klein2023,Pawbake2023}, even though the low temperature magnetic behavior is the one of a layered antiferromagnet~\cite{Wilson2021}. This low temperature behaviour was recently attributed to the ferromagnetic ordering of defaults~\cite{Long2023}. When magnetically ordered, bulk CrSBr is an ensemble of anisotropic FM planes, with an AFM coupling of adjacent planes~\cite{Avsar2022}. In multilayer CrSBr, the interlayer hopping is spin dependent which makes the energy of electronic states different for a AFM or FM ground state~\cite{Wilson2021}. This energy difference can be seen in the emission energy of the exciton which can then be used to determine for instance the value of saturation magnetic fields, and to characterize the magnetic ground state~\cite{Cenker2022}. The electronic properties of CrSBr are strongly anisotropic and this is reflected in its optical~\cite{Wilson2021,Klein2023a} and transport properties~\cite{Wu2022}. Low energy excitations in CrSBr include two magnon excitations at $25$ and $35$~GHz~\cite{Bae2022,Cham2022,Cho2023}. CrSBr exhibits two inequivalent magnetic ions in the magnetic unit cell and hence, a higher energy magnon branch~\cite{Bae2022}. Our infrared magneto-transmission experiments presented in Fig.~S1 of the Supplementary Materials, show an absorption at $372$~cm$^{-1}$ that disperses linearly with the external magnetic field with a g-factor close to 2, which we identify as the second magnon branch of bulk CrSBr.

To measure the low temperature magneto-optical properties of bulk CrSBr under hydrostatic pressure, we use the experimental set-up described in Ref.~\cite{Breslavetz2021}. A diamond anvil cell (DAC) is placed on piezo motors to allow for its displacement under the excitation laser spot produced by a long working distance objective. The pressure value at low temperature is determined by the photoluminescence of the ruby balls. Bulk CrSBr was grown by chemical vapor transport (see Supplementary Materials). Crystals with typical sizes of few tens of micrometers were inserted in the pressure chamber of the DAC as shown in Fig.~\ref{Fig1}b and could be oriented taking advantage of the strong shape anisotropy of this material that produces crystals elongated along the $\mathbf{a}$ crystallographic axis, see Fig.~\ref{Fig1}b. The DAC can be oriented at room temperature to apply the external magnetic field along different crystal directions (see Fig S2 of Supplementary Materials). This allows to compare the pressure dependent magneto-optical response giving access to the parameters describing the magnetic Hamiltonian. All measurements presented in this work have been performed at a temperature of $T=5$~K. Each value of the hydrostatic pressure was first applied at room temperature, and the DAC was then cooled down for optical spectroscopy measurements.

A typical low temperature photoluminescence spectrum of bulk CrSBr is presented in Fig.~\ref{Fig1}d. It includes an ensemble of sharp emission lines around $1.35$~eV related to excitons and exciton-polaritons~\cite{Wilson2021,Dirnberger2023}, while lower energy broad emissions are related to defects~\cite{Klein2023}.

\begin{figure}
\includegraphics[width=0.8\linewidth,angle=0,clip]{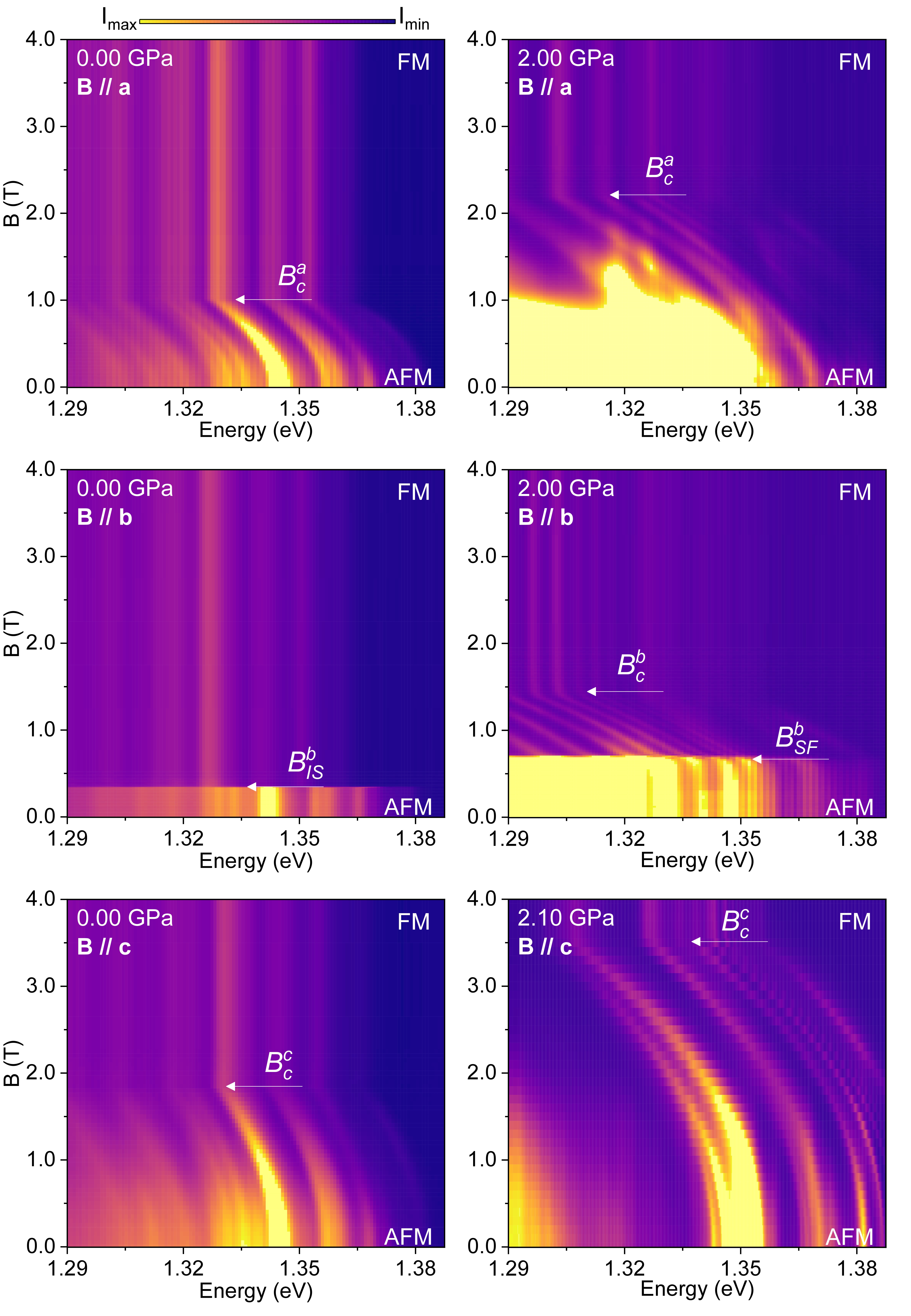}
\caption{False color maps of the low temperature magneto-photoluminescence response of bulk CrSBr for the  different values of hydrostatic pressure when the magnetic field is applied along the a axis (top row), the b axis (middle row) and the c axis (bottom row). The white arrows indicate the critical magnetic fields.
\label{Fig2}}
\end{figure}

\begin{figure}
\includegraphics[width=1\linewidth,angle=0,clip]{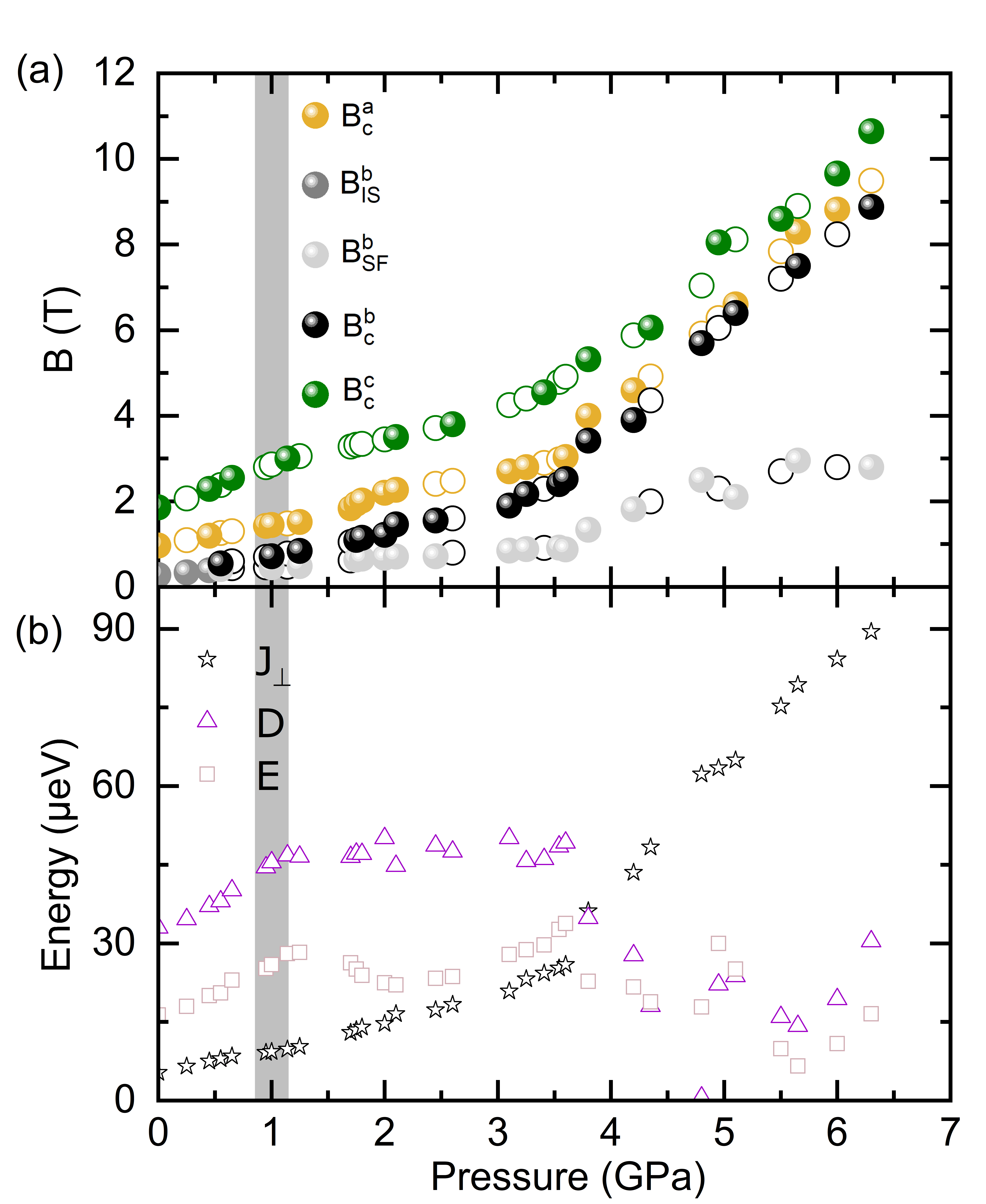}
\caption{a) Critical magnetic fields along the hard-c axis (green symbols), the intermediate-a axis (orange symbols) and along the easy-b axis (black symbols). Open symbols are extrapolated from nearby measurements. b) Evolution of the microscopic parameters describing the interlayer interaction $J_{\perp}$, the easy axis anisotropy $D$ and the intermediate-hard axis anisotropy $E$, as a function of the applied pressure. The spin-flop phase is observed when $B^b_{SF}< B <B^b_{c}$. The gray region indicates the range of pressure for which $2J_{\perp} \approx (D-E)$.
\label{Fig3}}
\end{figure}

We present in Fig.~\ref{Fig2} false color maps of the evolution of the low temperature photoluminescence spectrum of CrSBr at ambient pressure and at $P\approx 2$~GPa with the magnetic field applied along the $\textbf{a}$ (top row), $\textbf{b}$ (middle row) and $\textbf{c}$ axis (bottom row). When applying the magnetic field along the intermediate $\textbf{a}$ or the hard $\textbf{c}$ axis, the two spin sublattices gradually cant towards the direction of the external field. The exciton energy slowly evolves from the $B=0$ energy to a lower energy representative of the band structure of CrSBr with a FM magnetic ordering~\cite{Wilson2021}. When a hydrostatic pressure is applied, we observe a pronounced increase of the saturation magnetic fields $B^a_c$ and $B^c_c$ ($B^{\alpha}_c$ are the critical magnetic fields when the external magnetic field is applied along the $\mathbf{\alpha}$ axis), indicating a change in the competition between magnetic exchange and magnetocrystalline anisotropy. The complete evolution is presented in Fig. S3-S5 of the Supplementary materials. If the external $B$ field is applied along the $\textbf{b}$ axis, the observed behavior is different. At ambient pressure, the AFM to FM transition is abrupt (Ising transition or strong anisotropy regime) at a saturation field $B^b_{IS}$. When applying a hydrostatic pressure above $P=1$~GPa, we observe a different behavior: the exciton energy remains constant up to a saturation field $B^b_{SF}$ above which it initiates an approximately linear red-shift, up to a second value of the magnetic field labeled $B^b_{c}$. For $B > B^b_{c}$, the exciton emission energy remains constant, indicating that CrSBr is in a FM ground state. The intermediate behavior observed when $B^b_{SF} < B < B^b_{c}$ appears as a gradual decrease of the emission energy when increasing the magnetic field. This behavior is similar to that observed when the external magnetic field is applied along the $\mathbf{a}$ or $\mathbf{c}$ axis and describes the gradual canting of the spins towards the direction of the applied magnetic field.

To understand these results, we use the following microscopic spin Hamiltonian which is suitable for describing the two-sublattice orthorhombic AFM CrSBr:
\begin{equation}
\begin{split}
\hat{\mathcal{H}} = \sum_{\langle ij\rangle} J_{ij}\, \textbf{S}_{i} \cdot \textbf{S}_{j} &+
\sum_{i} \bigl\{ -D (S_i^{b})^2 + E\bigl[ (S_i^{c})^2- (S_i^{a})^2\bigr] \\
&- g_\alpha \mu_B B^\alpha S^\alpha_i\bigr\}.
\end{split}
\label{H0}
\end{equation}
Here ${\bf S}_i$ are $S=3/2$ spins of chromium ions, $J_{ij}$ are exchange coupling constants, $D$ and $E$ are parameters of biaxial single-ion anisotropy with $\alpha$ being either $\mathbf{b}$ (easy axis), $\mathbf{a}$ (intermediate axis ($D>E$)), and $\mathbf{c}$ the hard axis. Stable magnetic configurations can be determined by treating spins as classical vectors and minimizing the corresponding energy (\ref{H0}). In this way one can find that FM exchanges inside the $ab$ layers play no role in the magnetization process, which depends only on the AFM interlayer coupling $J_\perp > 0$. The three nearest neighbours exchange integrals have been determined by neutron scattering and their values are found to be two orders of magnitude larger than the interlayer one~\cite{Scheie2022}, so spins within the layers remain parallel to each other to minimize the intralayer exchange energy.

Magnetic field applied perpendicular to the easy-axis causes continuous spin tilting in the canted AFM state (CAF) till the full saturation is reached at the critical (second-order) field
\begin{eqnarray}
g_a \mu_B B^a_c & = & 2S(4J_\perp + D -E) \ ,\\
g_c \mu_B B^c_c & = & 2S(4J_\perp + D +E) \ .
\label{Hac}
\end{eqnarray}
where $g_{\alpha}$ are the g-factors when the external magnetic field is applied along the $\mathbf{\alpha}$ axis.

Once an external field is  applied along the easy axis of a collinear AFM, the magnetization remains zero till the spin-flop transition. The corresponding expression for the spin model of CrSBr
\begin{equation}
g_b \mu_B B^b_{\rm SF}  =  2S\sqrt{(D-E)(4J_\perp - D +E)}\ .
\label{Hbsf}
\end{equation}
Above $B^b_{\rm SF}$, the AFM sublattices rotate abruptly  towards the intermediate $a$ axis forming the CAF state. The full saturation is reached at
 \begin{equation}
g_b \mu_B B^b_c  = 2S(4J_\perp - D +E) \ .
\label{Hbs}
\end{equation}
Thus, the conventional magnetization process for a collinear AFM proceeds as
$\uparrow\downarrow \;\to\; {\rm CAF}\; \to\; \uparrow\uparrow$.

The above picture suggested by Louis N\'eel \cite{Neel1952}, needs to be modified for CrSBr. Because of a large  $D/J_\perp$, this AFM undergoes a direct first-order transition from the collinear AFM state into the fully polarized state typical for magnets with strong Ising anisotropy
\begin{equation}
g_b \mu_B B^b_{\rm IS}  = 4J_\perp S \ .
\label{His}
\end{equation}
The Ising-like transition for ${\bf B}\parallel{\bf b}$  has been experimentally observed in CrSBr~\cite{Goser1990,Wilson2021,Cho2023}. It corresponds to an abrupt reversal of magnetization in every second Cr-layer. Comparing the expressions (\ref{Hbsf})--(\ref{His}) we find that the conventional magnetization process is realized for $(D-E)<2J_{\perp}$, whereas for $(D-E)>2J_{\perp}$, the direct transition $\uparrow\downarrow \;\to\; \uparrow\uparrow$ takes place. The corresponding $T=0$ phase diagram is sketched in  Fig.~\ref{Fig1}c.

Our main experimental observation is the tuning of bulk CrSBr from the strong to the weak anisotropy regime for which a spin-flop phase is stabilized. Crossing this boundary is a unique possibility offered by bulk CrSBr for which $(D-E)$ and $2J_{\perp}$ are comparable at ambient pressure. The tuning is achieved by applying hydrostatic pressure which strongly enhances the interlayer coupling $J_\perp$ such that $(D-E)/2J_\perp$ becomes smaller. For $P>4$~GPa, the measured saturation field $B^b_{c}$ along the $\mathbf{b}$ axis increases and becomes almost equal to the one along the $\mathbf{a}$ axis. This evolution describes a decrease of the in-plane anisotropy ($(D-E)\rightarrow 0$) such that the ($\textbf{a},\textbf{b}$) plane becomes nearly isotropic.

Specifically, Fig.~\ref{Fig3}a shows the evolution of the measured values of the saturation magnetic fields along the three crystallographic axis. They strongly increase with the applied hydrostatic pressure and, at each pressure value, we can determine from Eq.~2-6 the three microscopic parameters $J_{\perp}$, $D$ and $E$. The evolution with pressure of these parameters is presented in Fig.~\ref{Fig3}b. At ambient pressure we find that $J_{\perp}=5.3~\mu$eV, $D=33.0~\mu$eV and $E=16.36~\mu$eV, in line with recent GHz investigations conducted on bulk material from the same batch~\cite{Cho2023}. At ambient pressure, bulk CrSBr is in the strong anisotropy regime fulfilling the condition $(D-E) > 2J_{\perp}$. When applying hydrostatic pressure, the lattice parameters and in particular the vdW gap, are changed and both the magnetic anisotropies and the interlayer exchange interaction are modified accordingly. $J_{\perp}$ is increased by $1700~\%$ at $P=6$~GPa, in line with recent results obtained at lower pressure values~\cite{Telford2023}. Both $D$ and $E$ also increase significantly, but in a similar way, keeping the difference $D-E$ at a constant value up to $P\sim 3$~GPa. The condition $(D-E) < 2J_{\perp}$, indicative of the weak anisotropy regime, is reached for $P \simeq 1$~GPa. Above this pressure value, we observe the intermediate spin-flop phase when applying an external magnetic field along the $\textbf{b}$ axis and two saturation fields as indicated in Fig.~\ref{Fig2} and \ref{Fig3}. The range of magnetic field within which the spin-flop phase is observed increases with the applied pressure, as the vdW gap shrinks and $J_{\perp}$ is increased. The variations of D or E are more likely related to deformations of the individual layers, along the c axis (bond angles) and within the plane of the layers, and also by a possible anisotropy of the Young modulus.

Above $P=3.5$~GPa, we note two main changes in the evolution of the microscopic parameters with pressure : Both D and E strongly decrease and become comparable, and the evolution rate of $J_{\perp}$ with pressure increases significantly. These two points are the sign of change in the system eventhough the observed behavior of the magneto-photoluminescence remains the same above and below this particular pressure value. As shown in Fig. S6 of the Supplementary Materials, applying hydrostatic pressure to bulk CrSBr is a reversible process.

To conclude, by applying high hydrostatic pressure to bulk CrSBr, we have modified the competition between the crystalline magnetic anisotropy and the interlayer exchange parameter. Based on low temperature magneto-photoluminescence measurements under such extreme experimental conditions, we have identified a spin-flop phase stabilized when the interlayer exchange interaction strongly exceeds the magnetic anisotropy. These results are understood in the frame of a spin Hamiltonian including magneto-crystalline anisotropies which describes the appearance of the spin-flop magnetic phase when $(D-E) < 2J_{\perp}$ (weak anisotropy regime). Compared to the experimental results, this model provides values for the microscopic parameters entering the spin-Hamiltonian and their evolution can be traced up to $P=6$~GPa. This work demonstrates the possibility of tailoring the competition between magnetic crystalline anisotropy and interlayer exchange interaction of vdW magnets by external means and enlarges the technical possibilities for the design of two dimensional magnets.

\begin{acknowledgements}

We thank Nicolas Ubrig and Gerard Martinez for valuable comments on the manuscript. This work has been partially supported by the EC Graphene Flagship project. M.E.Z. was supported by ANR Grant No. ANR-15-CE30-0004. M-A. M. acknowledges the support from the ERC (H2020) (Grant agreement No. 865826). N.P.W. acknowledges support from the Deutsche Forschungsgemeinschaft (DFG,German Research Foundation) under Germany’s Excellence Strategy—EXC-2111—390814868. We acknowledge the support of the LNCMI-EMFL, CNRS, Univ. Grenoble Alpes, INSA-T, UPS, Grenoble, France.

\end{acknowledgements}

%

\end{document}